\documentclass[]{article}  
\usepackage{graphicx}
\usepackage{amsfonts}
\usepackage{amssymb}
\usepackage{latexsym}
\usepackage{url}

\newenvironment{pf}{\unskip{\bf Proof:}}{\unskip{\hfill $\Box$}}

\newcommand{\lemlab}[1]{\label{lemma:#1}}
\newcommand{\theolab}[1]{\label{theo:#1}}

\newcommand{\eqlab}[1]{\label{eq:#1}}
\newcommand{\corlab}[1]{\label{cor:#1}}

\newcommand{\figlab}[1]{\label{fig:#1}}
\newcommand{\seclab}[1]{\label{section:#1}}

\newcommand{\lemref}[1]{\ref{lemma:#1}}

\newcommand{\theoref}[1]{\ref{theo:#1}}

\newcommand{\corref}[1]{\ref{cor:#1}}

\newcommand{\figref}[1]{\ref{fig:#1}}
\newcommand{\eqref}[1]{(\ref{eq:#1})}
\newcommand{\secref}[1]{\ref{section:#1}}

\newtheorem{theorem}{Theorem}
\newtheorem{lemma}[theorem]{Lemma}
\newtheorem{cor}[theorem]{Corollary}
\newtheorem{conj}{Conjecture}[section]

{\catcode`\@=11
\gdef\setft#1#2#3{%
\def\@oddfoot{
{\setbox0=\hbox{#1}
\setbox1=\hbox{#3}
\ifdim\wd0>\wd1
\dimen0=\wd0
\box0\hfil#2\hfil\hbox to\dimen0{\hfil\hfil\box1}
\else \dimen0=\wd1
\hbox to\dimen0{\box0\hfil }\hfil#2\hfil\box1 \fi
}}} }

\def\complaint#1{}
\def\withcomplaints{
\newcounter{mycomplaints}
\def\complaint##1{\refstepcounter{mycomplaints}%
\ifhmode%
\unskip%
{\dimen1=\baselineskip \divide\dimen1 by 2 %
\raise\dimen1\llap{\tiny -\themycomplaints-}}\fi%
\marginpar{\tiny [\themycomplaints]: ##1}}%
}

\def\a{{\alpha}}
\def\d{{\delta}}
\def\p{{\rho}}
\def\R{\mathbb{R}}
\def\mod{{\rm mod} \;}
\newcommand{\old}[1]{{}}

\newlength\abovesectionskip
\newlength\belowsectionskip
\makeatletter
\def\section{\@startsection{section}{1}{\z@}{-\abovesectionskip}%
               {\belowsectionskip}{\normalfont\Large\bfseries}}
\makeatother

\begin{document}

\title{{\bf Nonorthogonal Polyhedra\\Built from Rectangles}\thanks{%
Revision of the October 2001 version.
{\tt http://arXiv.org/abs/cs/0110059/}.%
}
}
\author{
Melody~Donoso
\and
Joseph~O'Rourke%
\thanks{
Dept. of Computer Science, Smith Col\-lege, North\-ampton,
MA 01063, USA.
\{mdonoso,\allowbreak orourke\}@\allowbreak cs.smith.edu.
Supported by NSF Distinguished Teaching Scholars award DUE-0123154.
}
}


\maketitle

\begin{abstract} 
We prove that any polyhedron of genus zero or genus one built out of
rectangular faces must be an orthogonal polyhedron, but
that there are nonorthogonal polyhedra of genus seven
all of whose faces are rectangles.
This leads to a resolution of a question posed by
Biedl, Lubiw, and Sun~\cite{bls-wcnfp-99}.
\end{abstract}
 
\section{Introduction}
\seclab{Introduction} 
A paper by Biedl, Lubiw, and Sun~\cite{bls-wcnfp-99}
raised the following intriguing question:
``Can an orthogonal net ever fold to a nonorthogonal polyhedron?''
We answer this question here: 
{\sc yes} for sufficiently high genus, and {\sc no} for sufficiently
small genus.
First we clarify the meaning of the terms in the question.

An \emph{orthogonal net} is an orthogonal polygon:  a simple, planar polygon
all of whose sides meet at right angles.
See, e.g., Fig.~\figref{Latin}.
\begin{figure}[htbp]
\centering
\includegraphics[width=0.25\linewidth,angle=90]{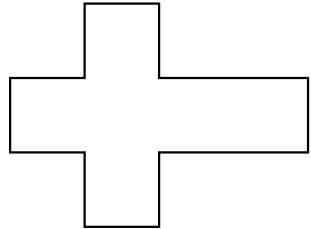}
\caption{A simple orthogonal polygon.}
\figlab{Latin}
\end{figure}
To \emph{fold} an orthogonal net means to crease it along
segments parallel to the polygon's sides and close up its perimeter.
Let us call this an \emph{orthogonal folding},
to contrast it with an arbitrary folding (when the creases can be
at arbitrary angles).
A \emph{polyhedron} is a surface formed of planar faces,
each a simple polygon, 
such that any pair is either disjoint, share one vertex, or
share an edge; such that each polyhedron edge is shared by exactly
two faces; and such that the ``link'' of each vertex is a cycle.
An \emph{orthogonal polyhedron} is a polyhedron all of whose faces
meet at edges with dihedral angles that are a multiple of $\pi/2$.
It is known that arbitrary foldings of orthogonal polygons can lead
to nonorthogonal polyhedra~\cite{lo-wcpfp-96}\cite{ddlop-mc-99}, 
but the question obtained
by restricting to orthogonal
foldings seems to be new.

We display their question, phrased in a positive sense, for later reference:

\begin{center}
\vspace{1mm}
\fbox{
\begin{minipage}[h]{0.75\linewidth}
{\bf Question~1}.
If a polyhedron is created by an
orthogonal folding of an orthogonal polygon, must it be
an orthogonal polyhedron?
\end{minipage}
} 
\vspace{1mm}
\end{center}

We first study a related question, which is divorced from the concept
of folding:

\begin{center}
\vspace{1mm}
\fbox{
\begin{minipage}[h]{0.75\linewidth}
{\bf Question~2}.
If a polyhedron's faces are all rectangles,
must it be an orthogonal polyhedron?
\end{minipage}
} 
\vspace{1mm}
\end{center}

We should note that the definition of
a polyhedron permits coplanar rectangles, so 
this second question could be phrased
as: ``if a polyhedron's faces can be partitioned
into rectangles,'' or, ``if a polyhedron can be constructed by gluing
together rectangles.''
Both Questions~1 and~2 can be viewed as seeking to learn whether,
loosely speaking,
orthogonality in $\R^2$ forces orthogonality in $\R^3$.

An orthogonal folding of an orthogonal polygon produces faces
that can be partitioned into rectangles.
Thus if rectangle faces force orthogonal dihedral angles,
then orthogonal foldings force orthogonal dihedral angles.
So we have the following implications:
\begin{itemize}
\item Q2: {\sc yes} $\Rightarrow$ Q1: {\sc yes}.
\item Q1: {\sc no} $\Rightarrow$ Q2: {\sc no}.
\end{itemize}
(The second is merely the contrapositive of the first.)
However, it is possible that the answer to Q2 is {\sc no} but
the answer to Q1 is {\sc yes}, for it could be that the 
nonorthogonal polyhedra made from rectangles cannot be unfolded
without overlap to an orthogonal net.  We first explore Q2 to remove
ourselves from the little-understood area of nonoverlapping unfoldings.
We will see that, nevertheless, we can ultimately prove the answer to both questions
is {\sc no} for polyhedra of
genus seven or above, but {\sc yes} for genus zero and one.

\section{A Nonorthogonal Polyhedron\\made with Rectangles}
\seclab{octopus}

Although this reverses the order of our actual development,
and might appear unmotivated, we first
present the example that shows that the answer to Question~2 is
{\sc no}.  The polyhedron is shown in two views in 
Fig.~\figref{octopus.1}.
\begin{figure}[htbp]
\begin{minipage}[b]{0.5\linewidth}
\centering
\includegraphics[width=0.9\linewidth]{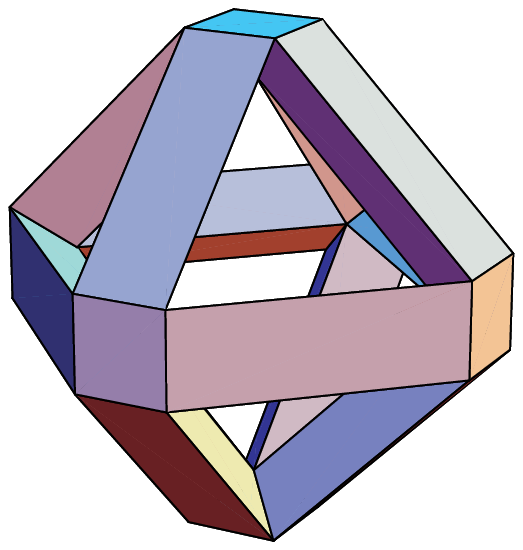}
\end{minipage}%
\begin{minipage}[b]{0.5\linewidth}
\centering
\includegraphics[width=0.9\linewidth]{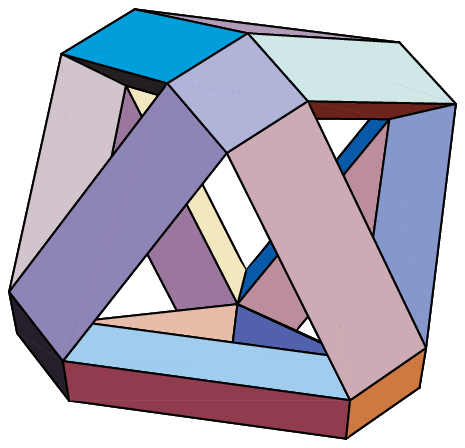}
\end{minipage}
\caption{An nonorthogonal polyhedron composed of rectangular faces.}
\figlab{octopus.1}
\end{figure}
It has the structure (and symmetry) of a regular octahedron,
with each octahedron vertex replaced by a cluster of five
vertices, and each octahedron edge replaced by a triangular
prism. Let us fix the squares at each octahedron vertex
to be unit squares.
The length $L$ of the prisms is not significant;
in the figure, $L=3$, and any length large enough to
keep the interior open would suffice as well.
The other side lengths of the prism are, however,
crucially important.
The open triangle hole at the
end of each prism has side lengths $1$ (to mesh with the
unit square), $\sqrt{3}/2$ and $\sqrt{3}/2$;
see Fig.~\figref{prism}.

This places the fifth vertex in the cluster displaced by $1/2$ 
perpendicularly from the center of each square,
forming a pyramid, as shown in
Fig.~\figref{pyramid}.

\begin{figure}[htbp]
\begin{minipage}[t]{0.45\linewidth}
\centering
\includegraphics[width=0.75\linewidth]{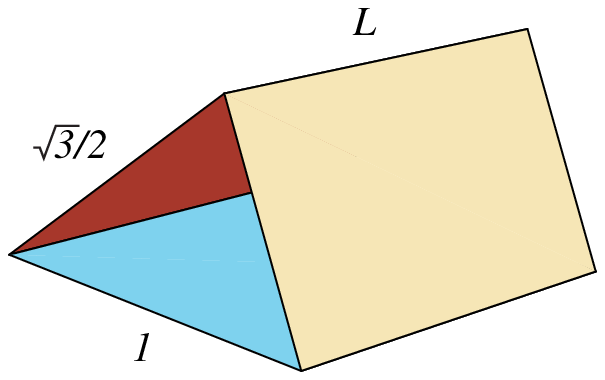}
\caption{Right triangular prism, used twelve times in Fig.~\figref{octopus.1}.}
\figlab{prism}
\end{minipage}%
\hspace{0.05\linewidth}%
\begin{minipage}[t]{0.5\linewidth}
\centering
\includegraphics[width=0.75\linewidth]{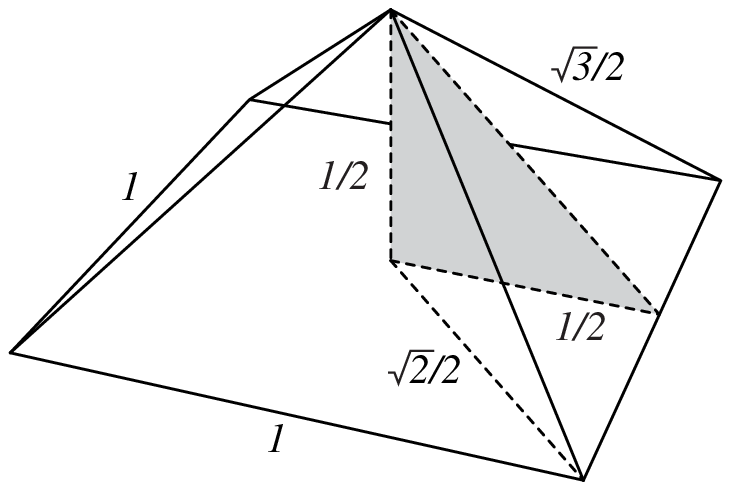}
\caption{The cluster of five vertices replacing each octahedron
vertex form a pyramid.}
\figlab{pyramid}
\end{minipage}
\end{figure}
The central isosceles right triangle (shaded in Fig.~\figref{pyramid}) 
guarantees
that two oppositely oriented incident triangular prisms
meet at right angles, just as do the corresponding edges of an octahedron.
The result is a closed polyhedron of
$V=6 \cdot 5 = 30$ vertices, $E=84$ edges, and $F=42$ faces.
The $42$ faces include $6$ unit squares,
$12$ $L \times 1$ rectangles,
and $24$ rectangles of size $L \times \sqrt{3}/2$.
Clearly the polyhedron is nonorthogonal: the three long
edges of each prism have dihedral angles of
approximately $54.7^\circ$
($=\tan^{-1} \sqrt{2}$),
$54.7^\circ$, and $70.5^\circ$ ($=\pi-2\tan^{-1} \sqrt{2}$),
two adjacent prisms meet at the same $70.5^\circ$ angle,
and the prisms meet the squares at $45^\circ$.
So not a single dihedral angle is a multiple of $\pi/2$.
 
By Euler's formula,
\begin{equation}
V -E+ F =  2(1-g)  \eqlab{Euler0}
\end{equation}
where $g$ is the genus.  From this 
we can compute the genus of the surface:
\begin{eqnarray*}
30 -84+ 42 = -12 & = & 2(1-g)\\
g & = & 7
\end{eqnarray*}
As it is easy to augment the polyhedron with attached 
(orthogonal) structures
that increase the genus,
this example establishes that the answer to Question~2 is {\sc no}:

\begin{theorem} 
There exist nonorthogonal polyhedra of genus $g \ge 7$
whose faces are all rectangles.
\theolab{g7}
\end{theorem}

Theorem~\theoref{g7} does not settle Question~1
immediately, for it could be that the polyhedron cannot
be unfolded to an ``orthogonal net.''
In fact, we were not successful in finding a nonoverlapping
unfolding of its surface.
However, it can be unfolded without overlap if boundary
sections are permitted to touch but not cross.
Such an unfolding could be cut out of a single sheet of paper,
but the cuts would have to be infinitely thin for perfect fidelity.
The unfolding is shown
in  Fig.~\figref{net.nonsimple}.
The solid edges are cuts; dashed edges are folds.
The folding is somewhat intricate, hinting at by the partial labeling
shown.
\begin{figure}[htbp]
\centering
\includegraphics[width=\linewidth]{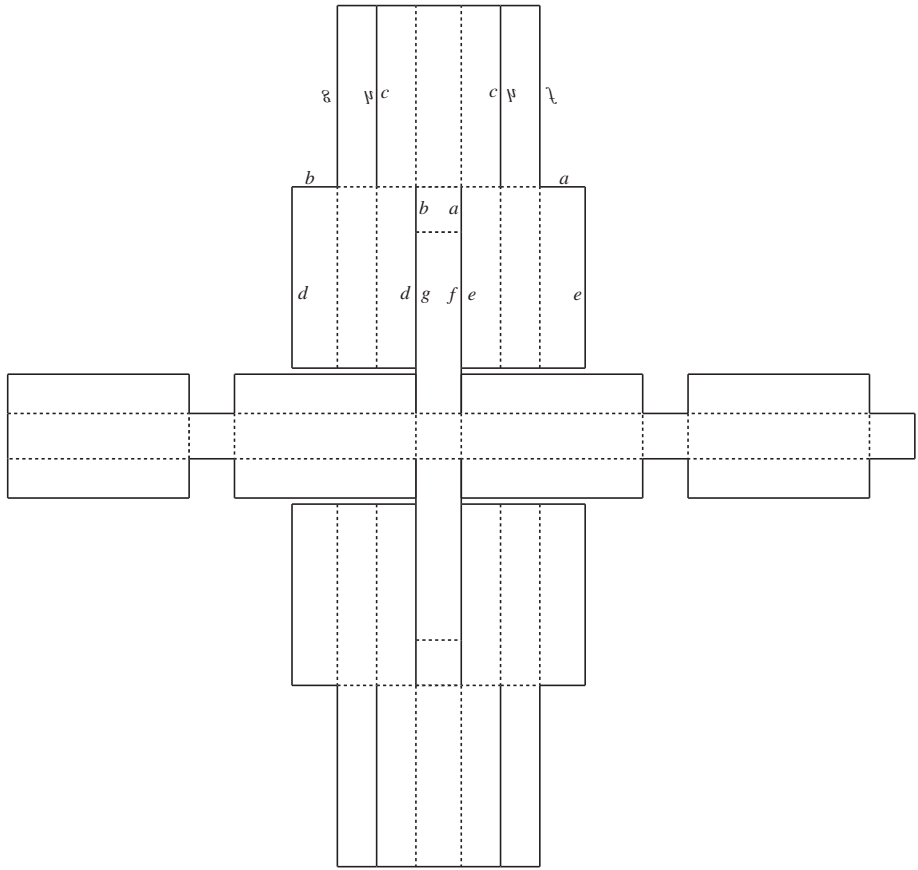}
\caption{A nonsimple net for the polyhedron of Fig.~\figref{octopus.1}.
The edges with the same label are glued to each other when folded.
Some labels are upside down, indicating the need for a spatial rotation
to reorient the edge before gluing.}
\figlab{net.nonsimple}
\end{figure}

However, the goal usually is to obtain a simple orthogonal polygon,
and such touchings violate the usual definition of a polygon.
A simple modification of the example 
illustrated in 
Fig.~\figref{octopus.3}
does lead to 
nonoverlap.
Each square is replaced by the five faces of a unit cube.
The polyhedron remains nonorthogonal (although now it has
some $\pi/2$ dihedral angles), and it may be unfolded to the
net shown in Fig.~\figref{oct}.
The role of the cubes is to spread out the unfoldings
of each prism so that they avoid overlap.

\begin{figure}[htbp]
\begin{minipage}[b]{0.5\linewidth}
\centering
\includegraphics[width=\linewidth]{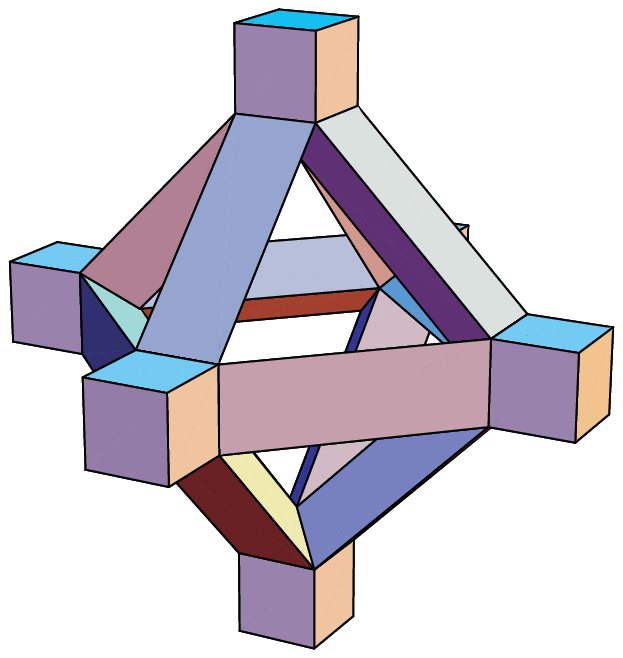}
\end{minipage}%
\begin{minipage}[b]{0.5\linewidth}
\centering
\includegraphics[width=\linewidth]{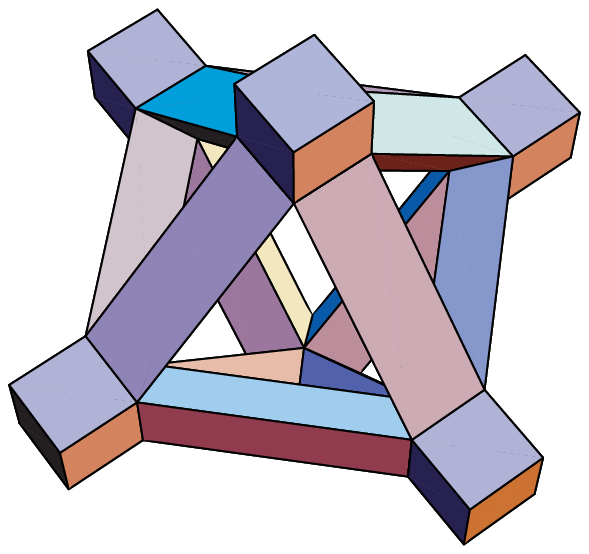}
\end{minipage}
\caption{Adding cubes to Fig.~\figref{octopus.1}.}
\figlab{octopus.3}
\vspace{2mm}
\centering
\includegraphics[width=0.9\linewidth]{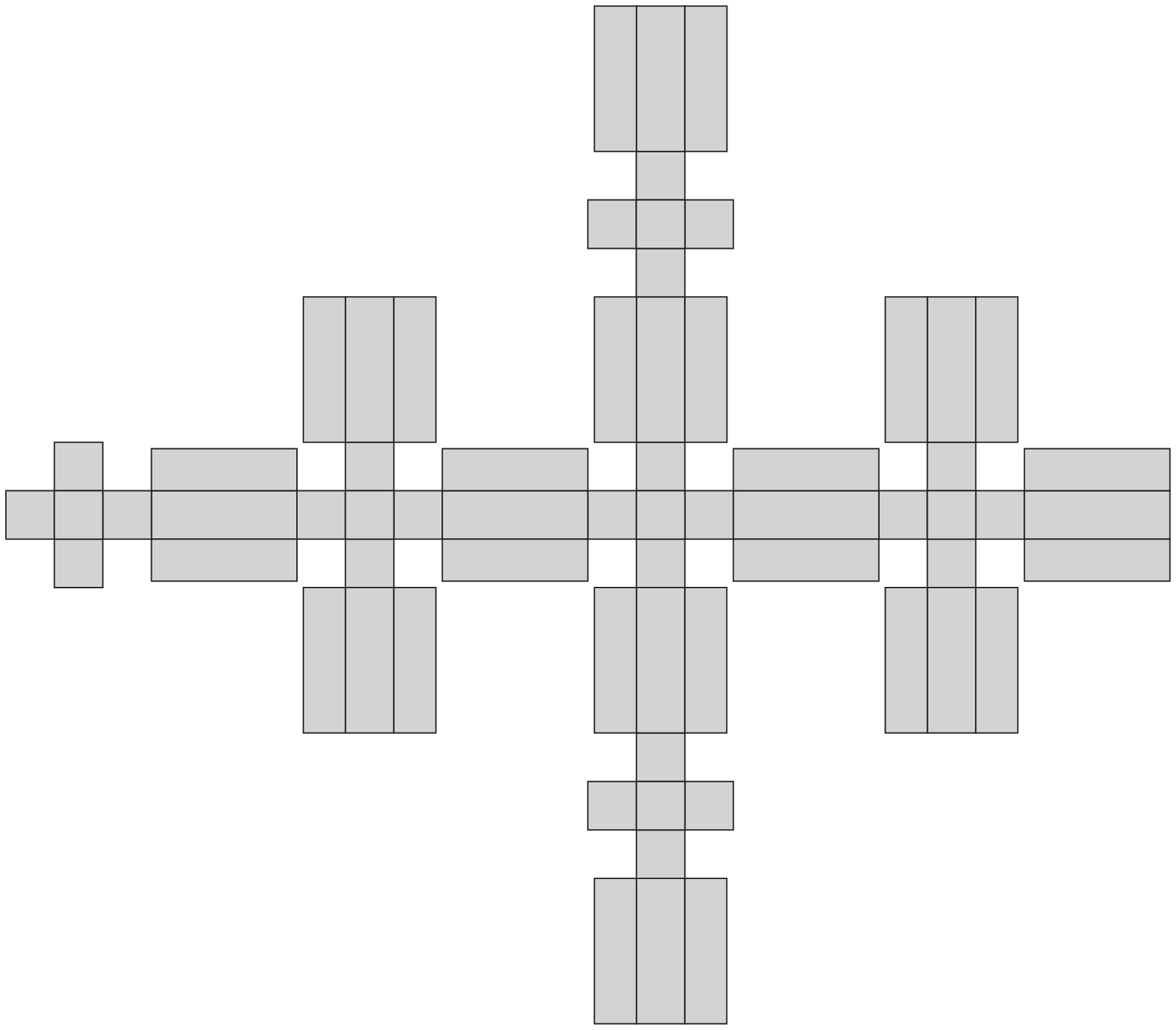}
\caption{An orthogonal polygon that folds to the nonorthogonal polyhedron
in Fig.~\figref{octopus.3}.}
\figlab{oct}
\end{figure}

This establishes that the answer to Question~1 is also {\sc no}:
\begin{theorem} 
There is a simple orthogonal polygon that folds orthogonally
to a nonorthogonal polyhedron.
\theolab{orthog.net}
\end{theorem}

\section{Overview of Proof}
\seclab{Overview}

The proof that the answer to Questions~1 and~2 
is {\sc yes} for sufficiently small genus
is long enough that a summary might be useful.
\begin{enumerate}
\item Sec.~\secref{Local}
establishes constraints on low-degree vertices, showing how
orthogonal and nonorthogonal dihedral angles may mix
locally at one vertex.
\item Sec.~\secref{Euler.Formula}
derives a corollary to Euler's formula (Eq.~\eqref{Euler0})
relating a lower bound on the number of edges
around a face to the average vertex degree.
\item Sec.~\secref{genus.0.1}
derives a lower bound of $4$ on the number of face edges
with nonorthogonal dihedral angle in a subgraph of
the $1$-skeleton of the polyhedron,
and uses this to establish the claim by contradiction.
\end{enumerate}

\section{Local Vertex Constraints}
\seclab{Local}

We now embark on a study of the possible configurations of
rectangles glued to one vertex of degree $\d$.
Each rectangle incident to a vertex glues
a face angle of $\pi/2$ there.  It is possible for
an arbitrary number $\d$ of such angles to be glued
to one vertex:  they can squeeze together like an accordian.
What is not possible is for there to be, in addition,
an arbitrary number
of dihedral angles that are multiples of $\pi/2$ incident
to one vertex.  It is this tradeoff we explore here.

\subsection{Conventions and Notation}

We need a term to indicate a dihedral angle that is a multiple
of $\pi/2$: such angles we call \emph{rectilinear},\footnote{
        The use of this term for this precise meaning is
        not standard.}
an angle $\a = k \pi/2$ for $k$ an integer.
For our purposes, the only rectilinear angles that will
occur are $\{ 0, \pi/2, \pi, 3\pi/2 \}$;
the $\pm$ sign of an angle will not be relevant.
An orthogonal polyhedron has only rectilinear dihedral angles.
The issue before us is:  Can there be a polyhedron, all of
whose faces are rectangles, which has at least one nonrectilinear
dihedral angle?
It will help to think of all edges with rectilinear dihedrals
as colored \emph{green} (good), and nonrectilinear dihedrals
as \emph{red} (bad).  We seek conditions under which a polyhedron
may have one or more red edges.

Our proofs will analyze the geometry around a vertex $v$
by intersecting the polyhedron $P$ with a small sphere
$S_v$ centered on $v$, and examining the intersections of the
faces incident to $v$ with $S_v$.
We normalize the radius of $S_v$ 
after intersection to $1$ so that an arc from pole to
pole has length $\pi$.
The intersection $P \cap S_v$ is a spherical polygon $P_v$ of
$\d$ great circle arcs, each $\pi/2$ in length
(because each face angle is $\pi/2$).
Edges incident to $v$ map to vertices of $P_v$,
and the dihedral angle $\a$ at an edge maps to the polygon
angle $\a$ on $S_v$ at the corresponding vertex.
$P_v$ is simple (non-self-intersecting) because the polyhedron
is simple.

With a Cartesian coordinate system centered on $v$, the intersection
of the $xy$-plane with $S_v$ is its \emph{equator},
the points of intersection of the $z$-axis the \emph{poles},
the six points at which the three axis pierce $S_v$ 
\emph{coordinate points},
and a great circle arc between coordinate points a
\emph{coordinate arc}.
Two points on $S_v$ are \emph{antipodal} if the
shortest arc between them has length $\pi$
(for example, the poles are antipodal).
We call two points a \emph{quarter pair} if
the
shortest arc between them has length $\pi/2$.
Note that neither antipodal points nor quarter pairs are
necessarily coordinate points.
Define the \emph{separation} $d(p_1,p_2)$ between two points
on $S_v$ to be the length of a shortest arc between them;
e.g., antipodal points have separation $\pi$.

Finally, define an \emph{orthogonal path} to be a simple path
$(p_0,p_1,\ldots,p_{n-1})$ on $S_v$, $n-1 \ge 1$,
such that adjacent points have separation $\pi/2$, and such that
the angle at all interior points, $p_1,\ldots,p_{n-2}$,
is rectilinear.
Note that a $\pi/2$-arc connecting two points is considered
an orthogonal path, with no interior points.
A consecutive series of rectlinear dihedral angles incident
to $v$ produce an orthogonal path on $S_v$.

\subsection{Preliminary Lemmas}

\begin{lemma}
An orthogonal path $\p=(p_0,\ldots,p_{n-1})$ with
one endpoint $p_0$ a coordinate point, and whose first arc
is a coordinate arc, must have the other endpoint
$p_{n-1}$ a coordinate point.
\lemlab{ortho.coord}
\end{lemma}
\begin{pf}
The proof is by induction on $n$. 
If $n-1=1$, the claim is trivial.
Otherwise, let 
$\p'=(p_1,\ldots,p_{n-1})$ 
be the path without the first arc. 
$\p'$ starts at a coordinate point (because
$p_0 p_1$ is a coordinate arc),
and its first arc is a coordinate arc (because the angle at $p_1$
is rectilinear).
Thus the induction hypothesis applies to $\p'$ establishing the claim.
\end{pf}

\begin{lemma}
If two points $p_1$ and $p_2$ are connected by an orthogonal path, they
are separated by a multiple of $\pi/2$,
and are therefore either antipodal or form a quarter pair.
\lemlab{antip.quart}
\end{lemma}
\begin{pf}
Rotate so that $p_1$ is a coordinate point,
and the arc of the path incident to $p_1$ is a coordinate arc.
Then
Lem.~\lemref{ortho.coord} applies and shows that $p_2$ is
a coordinate point.  Two coordinate points have a separation of
either $\pi$ or $\pi/2$.
\end{pf}

Of course this lemma implies (its contrapositive) that
two points $p_1$ and $p_2$ whose separation is not
a multiple of $\pi/2$ cannot be connected by an
orthogonal path.
\old{ 
\lemlab{ortho.not}
\end{lemma}
\begin{pf}
Suppose $p$ and $q$ could be connected by an
orthogonal path $\p$.  Rotate the coordinate system in $S_v$
so that $p_0$ is a coordinate point and the first arc of
$\p$ is a coordinate arc.  Then Lem.~\lemref{ortho.coord}
says that $q$ is a coordinate point.
But then the separation $d(p, q)$ must be a multiple of $\pi/2$,
a contradiction.
\end{pf}
}

\begin{lemma}
Let $(p_0,p_{n-1})$ be a quarter pair 
and let $a=p_0 p_{n-1}$ be the $\pi/2$-arc connecting them.
Every orthogonal path $\p$ between $p_0$ and $p_{n-1}$ forms
angles with $a$ which are both rectilinear,
(perhaps different) multiples of $\pi/2$.
\lemlab{quarter}
\end{lemma}
\begin{pf}
If $\p = a$ then the claim follows trivially with both angles $0$.
Suppose that $\p = (p_0,p_1,\ldots,p_{n-1}) \neq a$.
Rotate the coordinate system for $S_v$ so that $p_0$ is the north
pole, which places $p_{n-1}$ on the equator; 
see Fig.~\figref{quarter}.
\begin{figure}[htbp]
\centering
\includegraphics[width=0.375\linewidth]{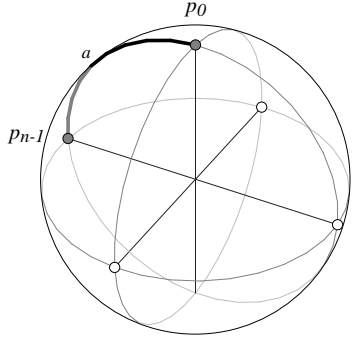}
\caption{$p_1$ must lie at one of the three remaining compass
points on the equator.}
\figlab{quarter}
\end{figure}
It must be that $p_1$ lies on the equator as well (because
each arc of an orthogonal path is of length $\pi/2$),
at a point different
from $p_{n-1}$ (because $\p \neq a$ and the path is simple).
Suppose that $d(p_1,p_{n-1})$ is not a multiple of $\pi/2$.
Then the contrapositive of
Lem.~\lemref{antip.quart} says there can be no orthogonal path
between them, contradicting the assumption that $\p$ is
an orthogonal path.
Thus $d(p_1,p_{n-1})$ must be a multiple of $\pi/2$.
This means that the $p_0 p_1$ arc makes an angle with $a$ that
is also a multiple of $\pi/2$.
Repeating the argument with the roles of $p_0$ and $p_{n-1}$
reversed leads to the same conclusion at the $p_{n-1}$ end.
\end{pf}

\old{  
Define the angle between two arcs that meet at $x$ on $S_v$ to be
the positive, smaller angle between them at $x$.

\begin{lemma}
Let $(p_0,p_{n-1})$ be an antipodal pair connected by an arc $a$.
Every orthogonal path $\p$ between $p_0$ and $p_{n-1}$ forms
angles with $a$ that are equal $\mod \pi/2$.
\lemlab{antipodal}
\end{lemma}
\begin{pf}
Rotate the coordinate system so that $p_0$ and $p_{n-1}$ are
the poles; then $p_1$ and $p_{n-2}$ are both on the equator.
The separation $s=d(p_1,p_{n-2})$ is 
$|\a_0-\a_{n-1}|$. 
see Fig.~\figref{antipodal}.
\begin{figure}[htbp]
\centering
\includegraphics[width=0.375\linewidth]{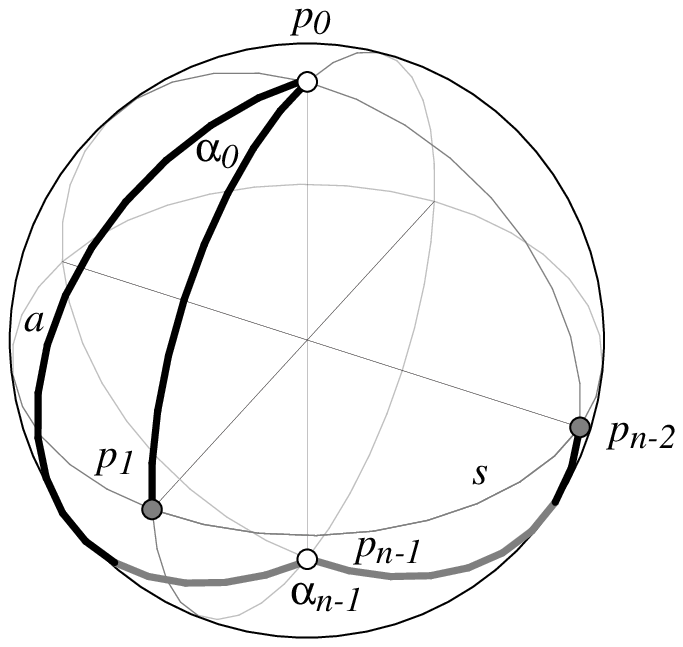}
\caption{$s = |\a_0-\a_{n-1}|$.}
\figlab{antipodal}
\end{figure}
Suppose, in contradiction to the lemma, that $|\a_0-\a_{n-1}|$ 
is not a multiple of $\pi/2$.  Then $s$ is not either.
Lem.~\lemref{ortho.not} then says that $p_1$ and $p_{n-2}$
cannot be connected by an orthogonal path.
But by assumption, $\p$ is an orthogonal path, a subpath of which
connects $p_1$ to $p_{n-2}$.  This contradiction
establishes the lemma.
\end{pf}
}

\subsection{Vertices with $k$ Red Edges}

We now employ the preceding lemmas to derive constraints
on vertices with few incident red edges
(edges at which the dihedral angle is not a multiple of $\pi/2$).
We will show that it is not possible to have just one, or
three incident red edges, that two and four red edges force
collinearities, and that zero or five (or more) red edges are
possible.  A summary is listed below.

\begin{center}\begin{tabular}{| c | l | c | }
        \hline
$k$
        & Constraint & Lemma
        \\ \hline \hline
$0$     &  possible & \mbox{}
        \\ \hline
$1$     &  not possible & Lem.~\lemref{1.red}
        \\ \hline
$2$     &  collinear & Lem.~\lemref{2.red}
        \\ \hline
$3$     &  not possible & Lem.~\lemref{3.red}
        \\ \hline
$4$     &  `{\tt +}' shape & Lem.~\lemref{4red:2+2}
        \\ \hline
$5$     &  possible & \mbox{}
        \\ \hline
\end{tabular}
\end{center}

\old{
\begin{lemma}
All three edges incident to a degree-3 vertex must have
dihedral angles of $\pi/2$, i.e., they must be green edges.
\lemlab{deg-3}
\end{lemma}
\begin{pf}
Let $P_v = (p_0,p_1,p_2)$.  Rotate so that $p_0$ is the
north pole.  Because both $p_1$ and $p_2$ are one $\pi/2$-arc
away from $p_0$, both lie on the equator.
Thus the arc $p_1 p_2$ is on the equator, and the dihedral
angles at $p_1$ and $p_2$ are $\pi/2$. The angle at $p_0$ is
also $\pi/2$, because it encompasses a quarter of the equator
spanned by $p_1 p_2$.
\end{pf}

\begin{lemma}
The dihedral angles of a vertex of degree-4 are
a cyclic permutation of $(\pi,\a,\pi,\a)$.
Thus it has at most two red edges.
\lemlab{deg-4}
\end{lemma}
\begin{pf}
[WRITTEN BUT NOT TYPED]
\end{pf}

\begin{cor}
If a degree-$4$ vertex has two red edges, they are collinear.
\end{cor}
\begin{pf}
[WRITTEN BUT NOT TYPED]
\end{pf}
}

Throughout the lemmas below, $e_i$ will represent the red edges,
and $p_i$ the corresponding points on $S_v$.

\begin{lemma}
No vertex has exactly one incident red edge.
\lemlab{1.red}
\end{lemma}
\begin{pf}
Suppose otherwise, and let $p_i$ be the corresponding
point of $P_v$ for red edge $e_i$.
Rotate the coordinate system so that $p_i$ is the north pole.
Then $p_{i-1}$ and $p_{i+1}$ both lie on the equator.
Now, $p_{i-1}$ is connected to $p_{i+1}$ by an orthogonal
path.
By Lem.~\lemref{antip.quart}, their separation must
be a multiple of $\pi/2$.
This in turn implies the angle at $p_i$ is rectilinear,
contradicting the assumption that $e_i$ is red.
\end{pf}

\begin{lemma}
If exactly two red edges are incident to a vertex,
they are collinear in $\R^3$.
\lemlab{2.red}
\end{lemma}
\begin{pf}
Let $p_i$ and $p_j$ be the points on $S_v$ corresponding
to the two red edges $e_i$ and $e_j$.
Because $p_i$ and $p_j$ are connected by an orthogonal
path, Lem.~\lemref{antip.quart} implies
they are antipodal or quarter pairs.
If they are antipodal, the claim of the lemma is
satisfied.
And indeed it is easy to realize this;
see Fig.~\figref{squares.4}.
\begin{figure}[htbp]
\centering
\includegraphics[width=0.5\linewidth]{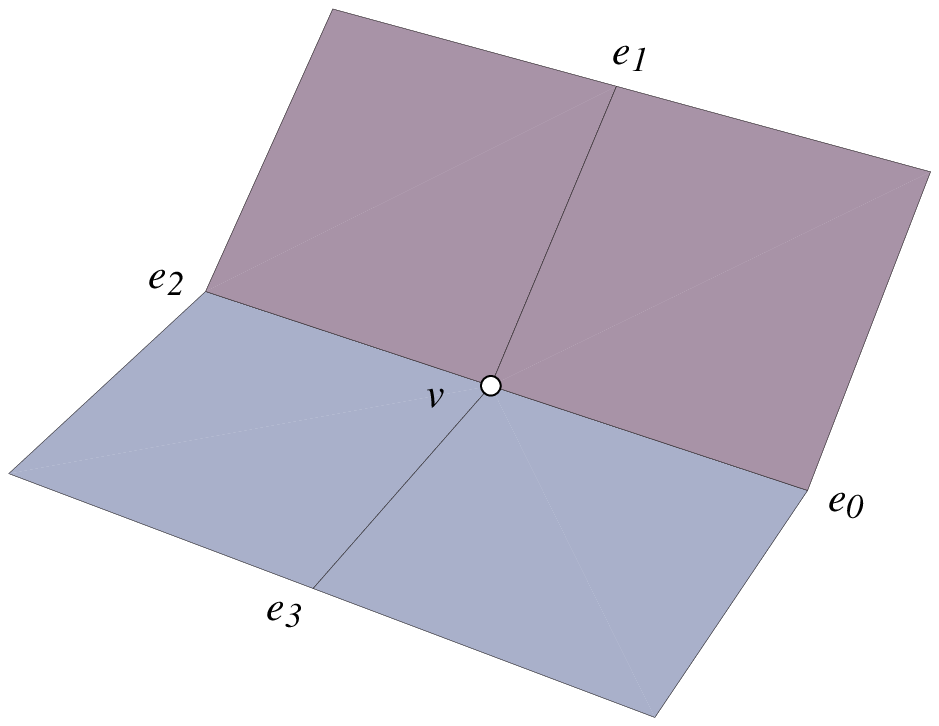}
\caption{Two red edges $e_0$ and $e_2$ incident to $v$.}
\figlab{squares.4}
\end{figure}
So suppose they form a quarter pair.
Let $a=p_i p_j$.
Then Lem.~\lemref{quarter} implies the
orthogonal path $(p_i,\ldots,p_j)$
makes a rectilinear angle at both $p_i$ and $p_j$.
The same holds true for the orthogonal path $(p_j,\ldots,p_i)$.
Therefore, $e_i$ and $e_j$ are green edges, a contradiction.
\end{pf}

Although we will not need this fact, it is not difficult to
prove
that the dihedral angles at the red edges in the previous
lemma are equal $\mod \pi/2$, as is
evident in Fig.~\figref{squares.4}.

\begin{lemma}
No vertex has exactly three incident red edges.
\lemlab{3.red}
\end{lemma}
\begin{pf}
Let the red edges be $e_i$, and $p_i$ their corresponding
points on $S_v$, $i=1, 2, 3$, 
and let $\p$ be the path on $S_v$.
Each of the consecutive pairs $(p_i, p_{i+1})$
either represents adjacent points of the path $\p$,
or points connected by an orthogonal path corresponding
to intervening green edges;
see Fig.~\figref{red.green}.
\begin{figure}[htbp]
\centering
\includegraphics[width=0.4\linewidth]{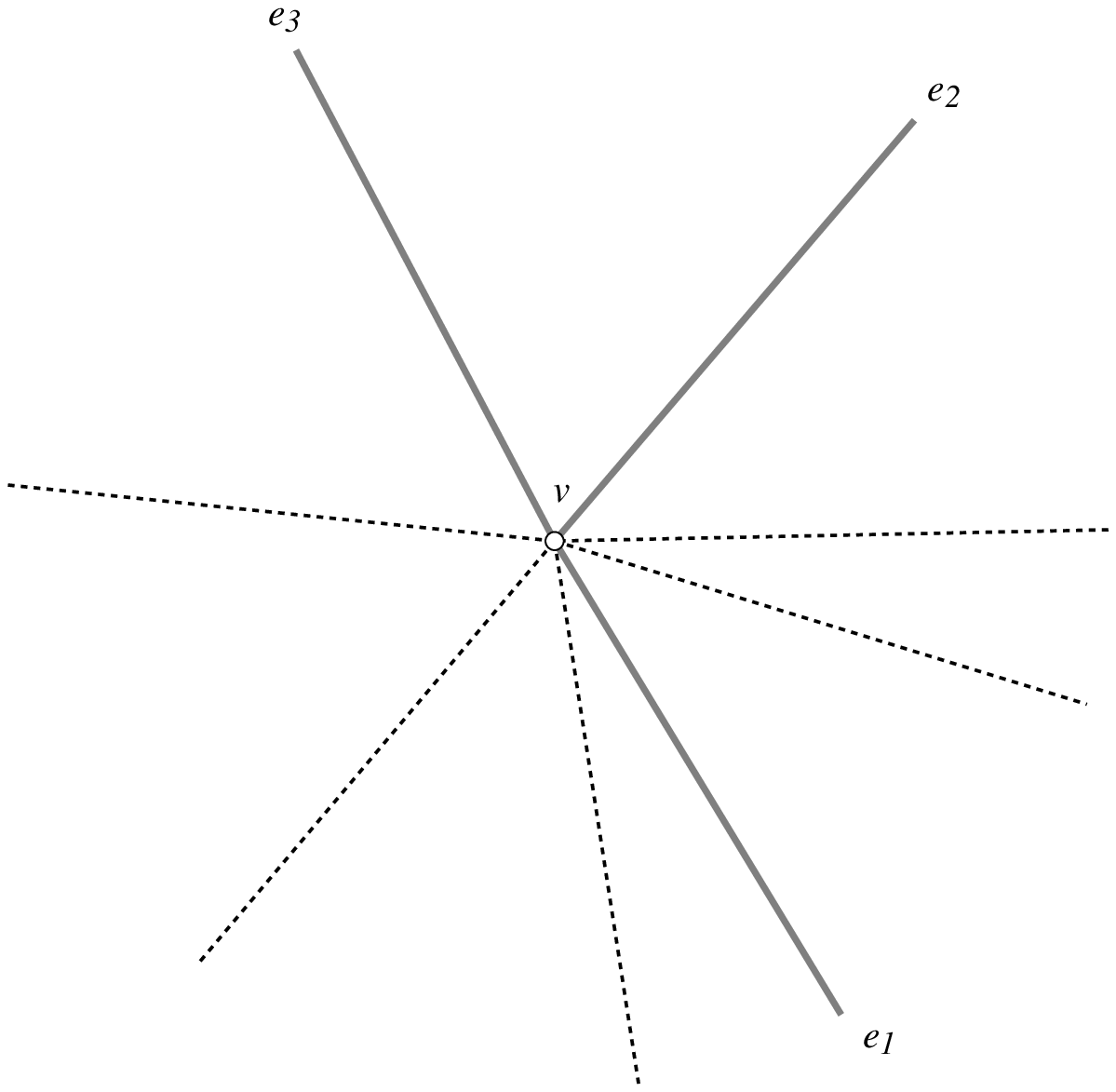}
\caption{Red edges $e_1$, $e_2$, $e_2$ separated by
green edges (dashed).}
\figlab{red.green}
\end{figure}

In the first case, we have a quarter pair;
in the second, Lem.~\lemref{quarter} implies that the
pair is either antipodal or a quarter pair.
We now consider cases corresponding to the number of
antipodal pairs.

\begin{enumerate}
\item Two or more pairs are antipodal.
This necessarily
places two points $p_i$, $p_j$, $i \neq j$, at the same location
on $S_v$, violating the simplicity of $\p$.

\item $(p_1,p_2)$ are antipodal, and $(p_2,p_3)$ and $(p_3,p_1)$ are quarter pairs.
Rotate so that $p_1$ and $p_2$ are poles;
then $p_3$ lies on the equator
(because it forms a quarter pair with both $p_1$ and $p_3$);
see Fig.~\figref{3.1}.
By Lem.~\lemref{quarter} the path $(p_2,\ldots,p_3)$ forms a rectilinear
angle with $a=p_2p_3$, as does
$(p_3,\ldots,p_1)$ with $a'=p_1p_3$.
Thus the angle of the path at $p_3$ must be rectilinear, and $e_3$ is a green edge,
a contradiction.

\begin{figure}[htbp]
\begin{minipage}[b]{0.5\linewidth}
\centering
\includegraphics[width=0.75\linewidth]{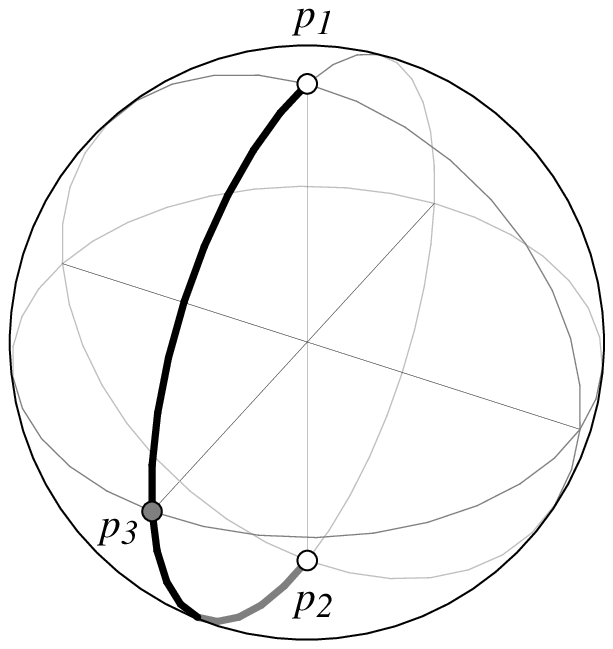}
\caption{Case 2.}
\figlab{3.1}
\end{minipage}%
\begin{minipage}[b]{0.5\linewidth}
\centering
\includegraphics[width=0.75\linewidth]{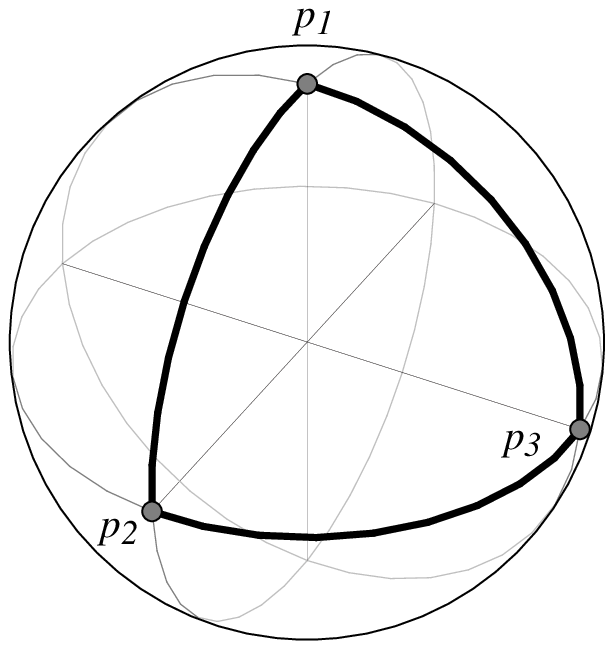}
\caption{Case 3.}
\figlab{3.2}
\end{minipage}
\end{figure}

\item All three are quarter pairs.
This forces the three points $p_1,p_2,p_3$ to lie at the corners
of a triangle with three $\pi/2$ angles; see Fig.~\figref{3.2}.
Applying Lem.~\lemref{quarter} to each of the connecting
orthogonal paths between these corners leads to a rectilinear
angle at all three points.  So all three edges $e_1,e_2,e_3$
are green, a contradiction.

\end{enumerate}
\end{pf}

\begin{lemma}
If a vertex has exactly four incident red edges, 
then they must fall into two collinear pairs
meeting orthogonally,
forming a `{\tt +}' in $3$-space.
\lemlab{4red:2+2}
\end{lemma}
\begin{pf}
Let the red edges be $e_i$, and $p_i$ their corresponding
points on $S_v$, $i=1, 2, 3, 4$, 
and let $\p$ be the path on $S_v$.
As in the proof of Lem.~\lemref{3.red},
each of the consecutive pairs $(p_i, p_{i+1})$ 
is either antipodal or a quarter pair.
We consider cases corresponding to the number of
antipodal pairs.

\begin{enumerate}

\item Three or four pair are antipodal.  This necessarily
places two points $p_i$, $p_j$, $i \neq j$, at the same location
on $S_v$, violating the simplicity of $\p$.

\item Two pair are antipodal.  Then all four points lie
on one great circle, say the equator.
Because each $p_i$ is connected to $p_{i+1}$ by an orthogonal
path, Lem.~\lemref{quarter} implies that their separation
must be a multiple of $\pi/2$.  To maintain simplicity, the four
points must be distributed at quarter-circle intervals.
This configuration is realizable, 
as is shown in Figs.~\figref{4.1.a} and~\figref{4.1.b}.
In both figures, $v$ is of degree $8$, but the
red/green pattern of incident edges is different.
The claim of the lemma is satisfied in that the red edges
form  a `{\tt +}'.
\begin{figure}[htbp]

\begin{minipage}[b]{0.5\linewidth}
\centering
\includegraphics[width=0.75\linewidth]{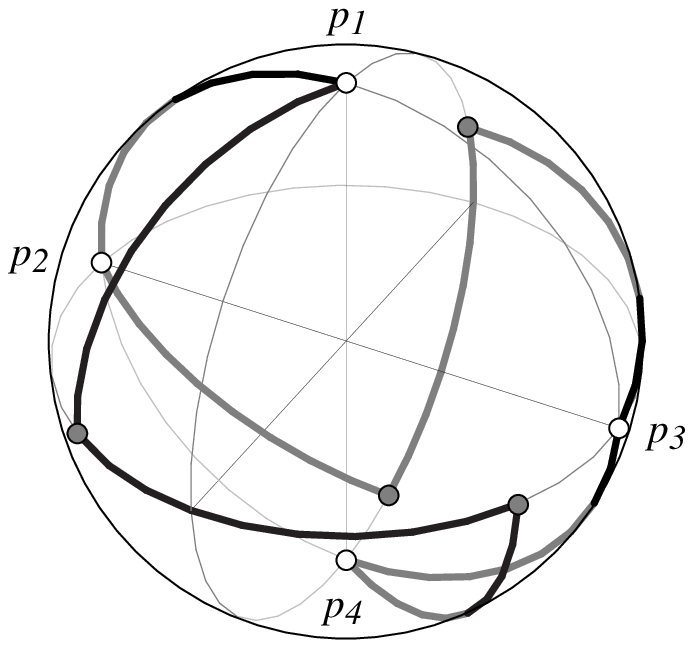}
\caption{Case 2:
$rrggrrgg$.}
\figlab{4.1.a}
\end{minipage}%
\begin{minipage}[b]{0.5\linewidth}
\centering
\includegraphics[width=0.75\linewidth]{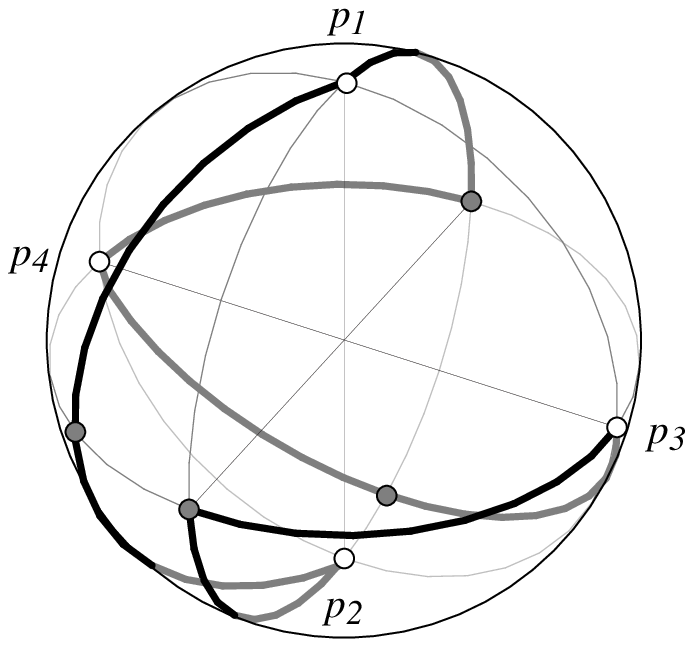}
\caption{Case 2:
$rgrgrgrg$.}
\figlab{4.1.b}
\end{minipage}

\end{figure}

\item One pair $(p_1, p_2)$ is antipodal, say at the poles,
and all others form quarter pairs.
Then $p_3$ and $p_4$ must lie on the equator,
separated by $\pi/2$;
see Fig.~\figref{4.2}.
The angle between the arcs $a=p_1p_4$ and $a'=p_4p_3$ 
(which are not necessarily part of the path $P_v$)
is $\pi/2$.
Applying Lem.~\lemref{quarter} to the orthogonal paths
connecting $p_1$ to $p_4$ and $p_4$ to $p_3$ leads
to a rectilinear angle in the path at $p_4$, a contradiction.

\item No pair is antipodal, so all are quarter pairs.
Rotate so that $p_1$ is the north pole;
then both $p_2$ and $p_4$ must be on the equator.

\begin{figure}[htbp]
\begin{minipage}[b]{0.5\linewidth}
\centering
\includegraphics[width=0.75\linewidth]{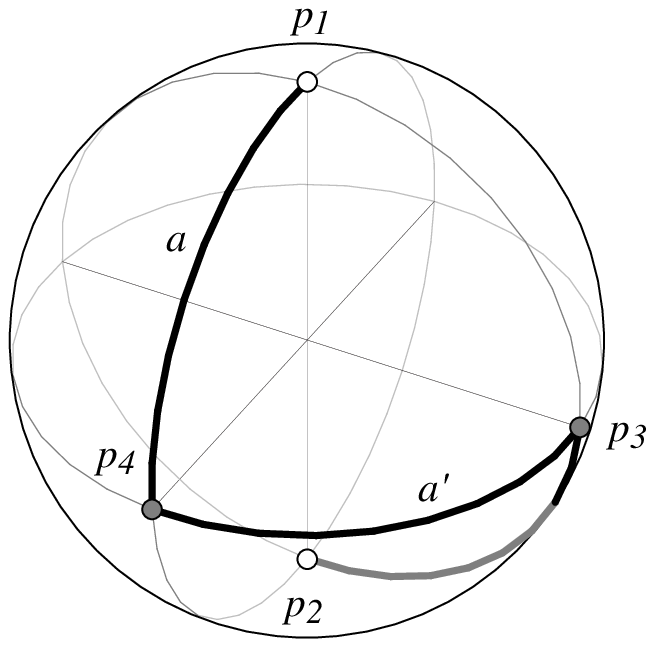}
\caption{Case 3.}
\figlab{4.2}
\end{minipage}%
\begin{minipage}[b]{0.5\linewidth}
\centering
\includegraphics[width=0.75\linewidth]{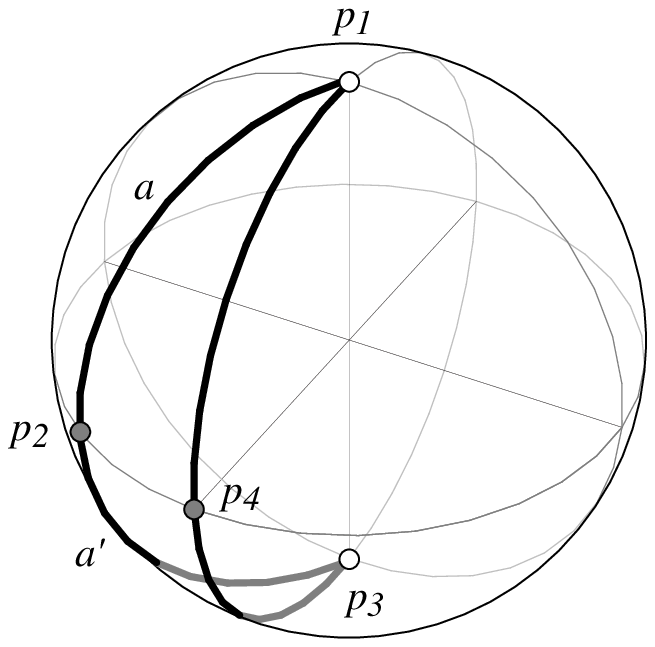}
\caption{Case 4.}
\figlab{4.4.b}
\end{minipage}
\end{figure}

\begin{enumerate}

\item $d(p_2,p_4)$ is a multiple of $\pi/2$.
Then the arcs $a=p_2 p_1$ and $a' = p_4 p_1$ meet at this
same rectilinear angle at $p_1$.  
Applying Lem.~\lemref{quarter} to the paths 
$(p_1,\ldots,p_2)$ and $(p_4,\ldots,p_1)$ leads to
a rectilinear path angle at $p_1$.
Thus $e_1$ is green, a contradiction.

\item $d(p_2,p_4)$ is not a multiple of $\pi/2$.
Then $p_3$ must lie on both the circle orthogonal to $p_2 v$
and the circle orthogonal to $p_4 v$.
These circles intersect at the poles, which forces $p_3$
to the south pole.
The arcs $a=p_1 p_2$ and $a' = p_2 p_3$ are necessarily
collinear at $p_2$; see Fig.~\figref{4.4.b}.
Applying Lem.~\lemref{quarter} to the paths 
$(p_1,\ldots,p_2)$ and $(p_2,\ldots, p_3)$
leads to a rectilinear angle at $p_2$, a contradiction.

\end{enumerate}
\end{enumerate}
\end{pf}

It is possible to have five red edges incident to a
vertex.
Fig.~\figref{deg5} shows one such example,
where a quarter pair $(p_1,p_5)$ is connected by
four arcs, connecting five edges none of whose dihedral angles
are rectilinear.
\begin{figure}[htbp]
\centering
\includegraphics[width=0.5\linewidth]{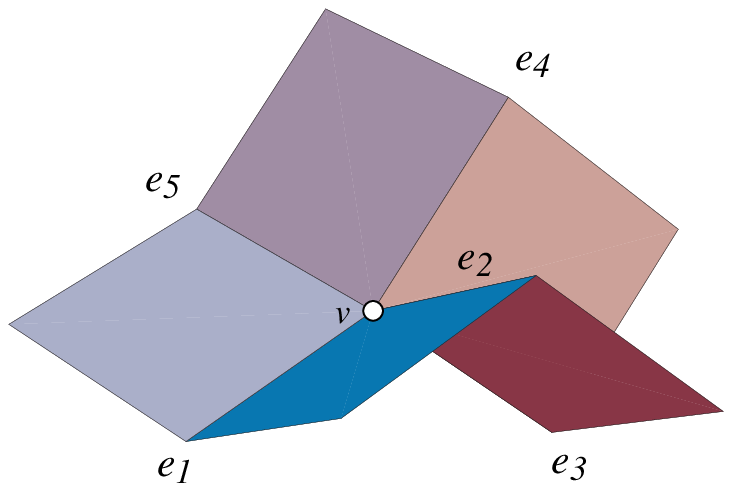}
\caption{Five red edges $e_1,\ldots,e_5$ incident to $v$.}
\figlab{deg5}
\end{figure}
It is also possible to have more than five red edges:
any larger number can accordian-fold into a small space,
say connecting a quarter pair of points.

The lemmas just derived will be employed in
Sec.~\secref{genus.0.1}.

\section{Euler Formula Computation}
\seclab{Euler.Formula}

Euler's formula says that
a (closed, bounded) polyhedron $P$ of
$V$ vertices, $E$ edges, and $F$ faces,
and of genus $g$, satisfies this linear relationship:
\begin{equation}
V -E+ F =  2-2g \;. \eqlab{Euler}
\end{equation}
The quantity $\chi=2-2g$ is known as the \emph{Euler characteristic}
of the surface.
Euler's formula applies to more general connected graphs:
those that are \emph{2-cell} embeddings, in that each of its faces
(the regions remaining when the vertices and edges are subtracted
from the surface) is a \emph{$2$-cell}, a region in which any simple closed
curve may be contracted to a point~\cite[p.~274]{co-aagt-93}.

We need a somewhat more general form of Euler's formula,
which applies to any connected graph embedded\footnote{
   To be \emph{embedded} means to be drawn without crossings.
} 
in a sphere with $h$ handles:
\cite[p.~84]{mt-gos-01} show that then
\begin{equation}
V - E + F  \ge  2 - 2 h \;. \eqlab{Euler.ineq}
\end{equation}
A sphere with $h$ handles is a surface of genus $g=h$.
This is more general for two reasons.
First,
not every such graph is the 1-skeleton of a polyhedron;
for example, a plane tree is 
in a surface of genus $0$, with one exterior 2-cell face.
Second, the faces might not be $2$-cells;
for example, a face could include a handle.

Our goal is to establish a corollary to Euler's formula
in the form of Eq.~\eqref{Euler.ineq}
expressed in terms of two quantities:
$d$, the average degree of a vertex,
and $k$, a lower bound on the number of edges in a boundary walk of a face.
The first quantity needs no explanation;
the second is straightforward for polyhedral graphs but needs
a definition for arbitrary graph embeddings.
A \emph{facial walk}~\cite[p.~100]{mt-gos-01}
visits all the edges bounding a face in a complete traversal.
The same edge might be visited twice in one facial walk,
as illustrated in Fig.~\figref{facial.walk}.
\begin{figure}[htbp]
\centering
\includegraphics[width=0.6\linewidth]{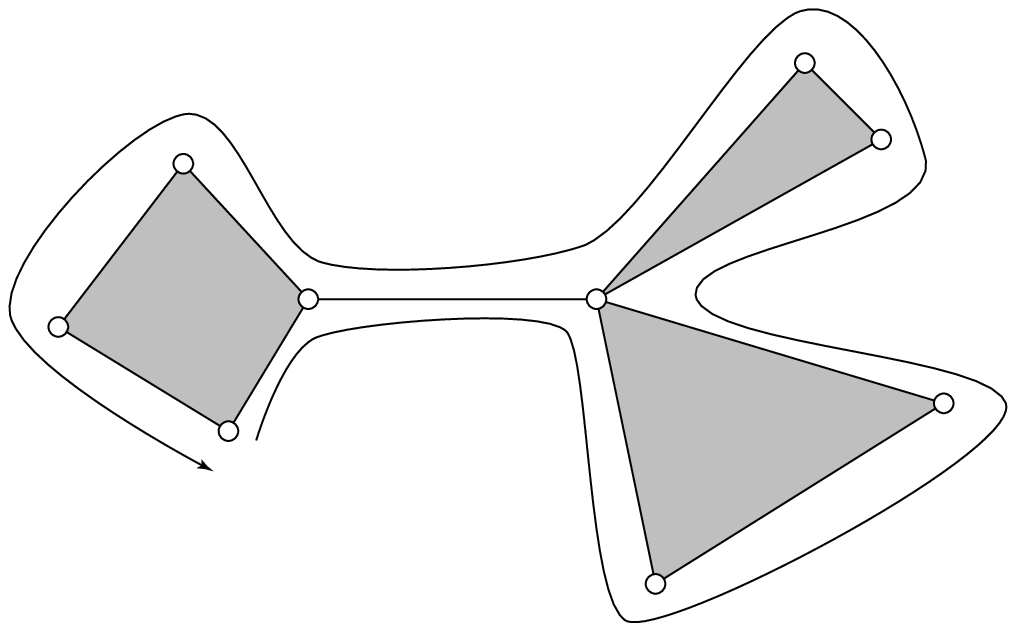}
\caption{The exterior face has $12$ edges in a boundary walk:
the $10$ surrounding the two triangles and quadrilateral,
and the central bridge edge counted twice.}
\figlab{facial.walk}
\end{figure}
If so, that edge is counted twice in its face-walk count.
The quantity $k$ reflects this possible double counting.
Each edge of $G$ either gets visited twice by one facial walk, 
or it appears in exactly two facial walks.
Each face of $G$ either contains one or more repeated vertices
or edges in its facial walk, or it is a cycle of $G$.

\begin{lemma}
Any embedding of a connected graph of $F$ faces on a surface
of genus $g$, whose average vertex degree is $d$ and
whose facial walks each include $k$ or more edges,
satisfies
\begin{equation}
F [ k - d(k-2)/2 ]  \ge  d \chi \;. \eqlab{chi.1}
\end{equation}
\lemlab{avg.deg}
\end{lemma}
\begin{pf}
The proof is an elementary counting argument, 
which, due to its unfamiliar form, 
we present in perhaps more detail than it deserves.
We start from the form of Euler's formula in Eq.~\eqref{Euler.ineq}:
\begin{eqnarray}
V - E + F  & \ge & \chi  \nonumber \\
V + F - \chi & \ge & E \eqlab{Euler.ineq2}
\end{eqnarray}

Because every edge is incident to two vertices, $d V = 2 E$,
and so:

\begin{equation}
V = \frac{2}{d} E \;.  \eqlab{V.E}
\end{equation}

Because every edge is counted exactly twice in facial walks,
and because $k$ is a lower bound on the number of edges in
any walk, we have $k F \le 2 E$, or

\begin{equation}
\frac{k}{2} F \le E \;.  \eqlab{F.E}
\end{equation}

Now we substitute Eq.~\eqref{V.E} into Eq.~\eqref{Euler.ineq2} to
eliminate $V$:

\begin{eqnarray}
\frac{2}{d} E + F - \chi & \ge & E \nonumber \\
F - \chi & \ge & E( 1 - 2/d ) \nonumber \\
\frac{F - \chi}{1-2/d} & \ge & E \;. \eqlab{left.half}
\end{eqnarray}

Putting this together with Eq.~\eqref{F.E} eliminates $E$:

\begin{eqnarray}
\frac{F - \chi}{1-2/d} & \ge & \frac{k}{2} F \nonumber \\
\frac{d (F - \chi)}{d-2} & \ge & \frac{k}{2} F \nonumber \\
d F - d \chi & \ge & (dk/2-k) F \nonumber \\
F [ d ( 1 - k/2 ) + k ] & \ge & d \chi \nonumber \\
F [ k - d(k-2)/2 ] & \ge & d \chi \;. \eqlab{chi.2} 
\end{eqnarray}
\end{pf}

To illustrate the import of this result, consider a 
polyhedron of genus zero:
$k=3$, because every face must
have at least three edges,
and $\chi = 2-2g = 2$.  Then Eq.~\eqref{chi.2} becomes
\begin{equation}
F [ 3 - d/2 ]  \ge  2d \;. \nonumber
\end{equation}
This requires the factor multiplying $F$ to be strictly positive:
$3-d/2 > 0$, which implies that $d < 6$.  This follows from the
familiar fact that the average vertex degree of a simple planar graph
is less than six.\footnote{
	E.g.,
	\cite[p.~104]{h-gt-72}, Cor.~11.1(e), or
	\cite[p.~446]{ah-dma-88}, Exer.~21.
}

In the next section we will apply Lem.~\lemref{avg.deg} in two circumstances:
genus zero and one, both with the lower bound $k=4$.
\begin{cor}
For connected graphs embedded on a surface of genus zero,
if $k = 4$, then $d < 4$.
\corlab{d.lt.4}
\end{cor}
\begin{pf}
Substituting $\chi=2$ and $k=4$ into Eq.~\eqref{chi.1} gives
\begin{equation}
F [ 4 - d ]  \ge  2d \nonumber
\end{equation}
which, in order for the factor of $F$ to be positive, implies that 
$d < 4$.
\end{pf}

\noindent
For example, a cube has $k=4$ and 
all vertices have degree $3$, so $d = 3 < 4$.
A more complex example is the 
``trapezoidal hexacontahedron,''\footnote{
	See, e.g., \url{www.georgehart.com} .
}
an Archimedian compound all of whose faces are quadrilaterals,
and so $k=4$.  It has $12$ degree-$5$ vertices, $30$ degree-$4$
vertices, and $20$ degree-$3$ vertices.
Thus its average vertex degree $d$ is 
$$ \frac{ 5 \cdot 12 + 4 \cdot 30 + 3 \cdot 20}{12+30+20} = \frac{240}{62} \approx 3.87 \;.$$

\begin{cor}
For connected graphs embedded on a surface of genus one,
if $k = 4$, then $d \le 4$.
\corlab{d.le.4}
\end{cor}
\begin{pf}
Substituting $\chi=0$ and $k=4$ into Eq.~\eqref{chi.1} gives
\begin{equation}
F [ 4 - d ]  \ge  0 \nonumber
\end{equation}
Now it could be that the factor of $F$ is zero;
so we must have $4-d \ge 0$, i.e.,
$d \le 4$.
\end{pf}

\noindent
For example, a cube with a rectangular hole connecting top and bottom
faces leads to $k=4$ when the punctured 
top and bottom faces are partitioned
into four quadrilaterals each.
All vertices then have degree $4$, so $d = 4$.

\section{Orthogonality Forced for Genus Zero and One}
\seclab{genus.0.1}

We establish in this section that the answer to Question~2 is {\sc yes}
for polyhedra of genus zero and one.
We first define a red subgraph, then prove that $k=4$ for it,
and, finally, exploit the two corollaries above.

\subsection{Red Subgraph $G_r$}
Starting with the 1-skeleton $G$ of the polyhedron, 
with its edges colored green or red depending on whether
the dihedral angle is rectilinear or not, we
perform the following operations to reach $G_r$:
\begin{enumerate}
\item Remove all green edges from $G$, retaining only red edges.
\item Merge edges meeting at a degree-$2$ vertex:
if a node $y$ is of degree $2$, with incident edges $(x,y)$ and $(y,z)$,
delete $y$ and those edges and replace with $(x,z)$.
\item Select one component of the resulting graph and call it
$G_r$, the \emph{red subgraph}.
\end{enumerate}
If $G$ contains any red edge, then there is a nonempty $G_r$.
Note that $G_r$ is realized in $\R^3$ with straight segments
for each edge, because
Lem.~\lemref{2.red} guarantees that the ``erasing'' of degree-$2$
vertices merges collinear polyhedron edges.

\subsection{Facial Walks in $G_r$}

$G_r$ is naturally embedded on the surface of the polyhedron.
Call this its \emph{canonical embedding}.

\begin{lemma}
Every facial walk in the canonical embedding of $G_r$ 
contains at least four edges.
\lemlab{cycle.4}
\end{lemma}
\begin{pf}
Let $F$ be a face in the canonical embedding of $G_r$,
and $W=(e_1,e_2,\ldots,e_m)$ its face walk.
We will show that each $m \le 3$ leads to a contradiction.
\begin{enumerate}
\item $W$ contains just one edge.
Then the edge is a loop; but $G_r$ is loopless.
\item $W$ contains just two edges.
Then it must walk around a ``dangling'' red edge.
Then the vertex $v$ at the end of this edge must be degree 1 in $G_r$,
in contradiction to Lem.~\lemref{1.red}.
\item $W$ contains just three edges.
In $G$, $e_i$ and $e_{i+1}$
are separated by green edges (or no edges).
Let $v$ be the vertex shared by $e_i$ and $e_{i+1}$.
Then $p_i$ and $p_{i+1}$ are connected by an orthogonal
path on $S_v$.
Lem.~\lemref{antip.quart} then says their separation on $S_v$
is a multiple of $\pi/2$.
Thus, in $\R^3$, the geometric angle between $e_i$ and $e_{i+1}$
is $\pm \pi/2$, or it is $\pi$--i.e., they are collinear.
Suppose at least one vertex has angle $\pi$, so that, say,
$e_1$ and $e_2$ are collinear.  Then $e_3$ must be collinear as
well, and we necessarily have edges overlapping collinearly
in $\R^3$.  But all three edges are distinct nonoverlapping
segments on the polyhedron surface, so this is a contradiction.
Suppose, then, that all three vertices have angle $\pm \pi/2$.
Fixing attention on $e_2$, $e_1$ and $e_3$ lie in parallel planes
perpendicular to $e_2$ and through its endpoints.  
Thus, $e_1$ and $e_3$ cannot
close to a triangle in $\R^3$, again a contradiction.
\end{enumerate}
\end{pf}

\noindent
Four edges are needed to close a cycle in $\R^3$;
four right angles force the cycle edges to lie in a plane,
and therefore form a rectangle.

\subsection{Concluding Theorem}

\begin{theorem}
Any polyhedron of genus zero or one, all of whose faces are rectangles,
must be an orthogonal polyhedron: all of its dihedral
angles are multiples of $\pi/2$.
\theolab{g01}
\end{theorem}
\begin{pf}
By Lem.~\lemref{cycle.4} we know that 
every face walk
of $G_r$ contains at least four edges.  Thus
$k=4$ in the notation of Lem.~\lemref{avg.deg}.
\begin{description}
\item[$g=0$]
Applying Cor.~\corref{d.lt.4} leads to the conclusion
that $d$ must be strictly less than $4$, which implies
that $G_r$ must include at least one vertex of degree $3$.
We defined $G_r$ to merge all degree-2 nodes, so it has none of those.
Lem.~\lemref{1.red} and~\lemref{3.red} show that it can have no nodes
of degree 1 or 3 respectively.
Thus the minimum degree of a node of $G_r$ is $4$,
and an average degree $d < 4$ is impossible.
\item[$g=1$]
Applying Cor.~\corref{d.le.4} leads to the conclusion
that $d \le 4$.  As above, the minimum degree of a node
of $G_r$ is $4$.
So then we must have every node of $G_r$ exactly degree $4$.
(This is precisely what is achieved in the cube-with-a-hole example,
incidentally.)
Now Lem.~\lemref{4red:2+2} says that the edges
incident to a degree-4 vertex of
$G_r$ come in two collinear pairs.
We now argue that this implies that the surface contains
an infinite line, a contradiction to the fact that a polyhedron
is bounded.

Let $e_1$ be an edge incident to a vertex $v_1$ of $G_r$.
$v_1$ must be of degree 4 as above, and therefore Lem.~\lemref{4red:2+2}
provides an $e_2$ collinear with $e_1$.  Call the other endpoint
of $e_2$ vertex $v_2$.  Repeating the argument leads to an edge $e_3$
collinear with $e_2$.  In this way we produce a sequence of
collinear edges, $e_1, e_2, \ldots$.  There is no end to this
sequence, providing a contradiction.
\end{description}
\end{pf}

One way to view the above proof is that a red cycle cannot turn
in $\R^3$ with only vertices of red-degree no more than 4.
It requires degree-5 vertices, or higher degree
vertices, to permit a cycle to form.
This is exactly how the polyhedron described in Sec.~\secref{octopus}
is constructed.\footnote{
	This is, in fact, how it was discovered.}
The four corners of each square in 
Fig.~\figref{octopus.1} are red-degree $5$ vertices, and the 
fifth vertex 
$\frac{1}{2}$ offset from the center of each square is red-degree $8$,
as illustrated in 
Fig.~\figref{oct.cutopen}.
\begin{figure}[htbp]
\centering
\includegraphics[width=0.4\linewidth]{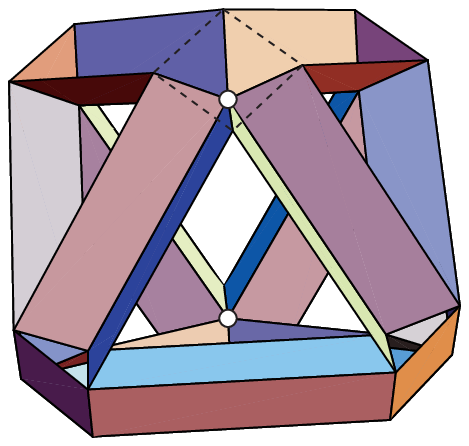}
\caption{The polyhedron in
Fig.~\figref{octopus.1} with five top faces
removed. The indicated vertices have degree $8$ in $G_r$:
$8$ incident nonrectilinear edges.}
\figlab{oct.cutopen}
\end{figure}

\subsection{Genus Two and Higher}
\seclab{genus.ge.2}
It seems the computations used in the previous
theorem provide no useful constraint
when $g \ge 2$.  
Then $\chi \le -2$, and Eq.~\eqref{chi.1}
only yields (for $k=4$)
\begin{eqnarray*}
F ( 4 - d ) & \ge & -2\\
F & \ge & \frac{2}{d-4}
\end{eqnarray*}
which can be satisfied for all $d$.
So there is effectively no constraint on $d$ for $g \ge 2$.
This seems to reveal the limit of this proof technique.

\section{Discussion}
\seclab{Discussion}
The obvious open problem is to close the 
gap---between
Theorem~\theoref{g01},
$g \le 1$, when orthogonality
in $\R^3$ is forced, and 
Theorem~\theoref{g7},
$g \ge 7$, when it is not forced.
Is there a nonorthogonal polyhedron of genus $g$,
$2 \le g \le 6$, constructed entirely from rectangles?

Perhaps a more interesting 
problem is to extend these results to other planar constraints,
and ask if they imply a restriction in $\R^3$.
For example, if a polyhedron is constructed out of
convex polygons whose angles are all
multiples of $\pi/k$, $k > 2$, 
does this imply any restriction
on the realizable dihedral angles?

Finally, because so little is known about nonoverlapping (simple)
edge-unfoldings, perhaps just-barely nonsimple unfoldings,
of the type exemplified by Fig.~\figref{net.nonsimple}, 
should be considered.

\paragraph{Acknowledgments}
We thank Michael Albertson for advice on graph embeddings,
Erik Demaine for useful comments,
Paul Nijjar for finding an error in an earlier version of
Lemma~13, and Timothy Chan for finding an error in a (now removed)
figure.

\bibliographystyle{alpha}
\bibliography{/home1/orourke/bib/geom/geom}

\end{document}